# The spherical coordinate form of three-dimensional generalized dynamics of soft-matter quasicrystals with 12-fold symmetry

Zhi–Yi Tang [1]*, Tian-You Fan [2]

1 School of Computer Science and Technology, Beijing Institute of Technology, Beijing 100081, China
2 School of Physics, Beijing Institute of Technology, Beijing 100081, China
*tangtangd929@sina.com

**Abstract** This article reports the spherical coordinate form of three-dimensional generalized dynamics of soft-matter quasicrystals with 12-fold symmetry which provides a basis for solving initial-boundary value problems of the equations under some important cases. Some relevant solving methods are discussed as well.

**1. Introduction**

Soft-matter quasicrystals of 12- and 18-fold symmetries were observed extensively in liquid crystals, polymers, colloids, nanoparticles and surfactants and so on [1-12]. This is an important event in 21th century chemistry. The soft-matter quasicrystals are formed through self-assembly of spherical building blocks by supramolecules, compounds and block copolymers and so on, which is associated with chemical process and is quite different from that of solid quasicrystals. The new structure presents both natures of soft matter and quasicrystals. Soft matter is an intermediate phase between ideal solid and simple fluid, which exhibits fluidity as well as complexity as pointed out by de Gennes[13], while quasicrystals are highly ordered phase. We can say the soft-matter quasicrystals are complex fluid with quasiperiodic symmetry. Refs [14-24] reviewed soft-matter quasicrystals from different angles on their formation mechanism, structure stability, thermodynamics and the correlation between Frank-Kasper phase and quasicrystals etc. The dynamics of soft-matter quasicrystals is an important subject of the study. To describe the fluidity Fan and his group[25-28] introduced a new elementary excitation--- fluid phonon besides the phonon and phason as described above for solid quasicrystal study, in which the concept of fluid phonon was originated from Landau school [29]. Due to the existence of the fluid phonons, equation of state is necessary; Fan and Fan [30] modified the equation of state for dense columnar liquid crystals suggested by Wensink [31], so the governing equations of a generalized dynamics of the matter are set up. With the equations Li and Fan [32] found the solutions of dislocation for the new phase, Cheng et al solved the flow past circular cylinder and impact tensile specimen for 5- and 10-fold symmetry soft-matter quasicrystasls refer to [33,34], Cheng et al [35] studied the crack and rupture problem of the matter, Wang et al [36, 37] reported the solutions of the transient response problems of soft-matter quasicrystals of 8- and 14-fold symmetries. In addition, Fan and Tang [38], Tang and Fan[39] studied the thermodynamics concerning the generalized dynamics, and obtained the stability criteria of the soft-matter quasicrystals. These solutions examined the equations and formulations, explored the structure and properties of the matter, and show the meaning of the generalized dynamics. However the discussions and solutions reported in Refs [32-39] are limited in the range in two dimensions and in the form of rectilinear coordinate system, present their limitation. There are lots of needs for solving of three-dimensional problems, for example, flow of soft matter past a sphere is an important topic. In the conventional fluid dynamics, there was the famous Stokes solution, which plays an important role in modern physics[40]. Einstein made use of it in the theory of the Brownian motion which leads to a determination of Loschmidt's number. Later it has become important again in Millikan's determination of the electronic charge. In the studies of soft-matter quasicrystals the sphere problems are also important; refer to Yue et al [12] and Huang et al [15]. The generalized Stokes problem of flow of soft-matter quasicrystals is significant too, and the motion of sphere and the stress analysis have been carrying out. For the purpose we must developprobe on spherical coordinate form of the three-dimensional dynamic equations.

## 2. Fundamental of generalized dynamics of soft-matter quasicrystals

The generalized dynamics of soft-matter quasicrystals was suggested in Refs [25-28], the general form of the governing equations are as below

$$\frac{\partial \rho}{\partial t} + \nabla_k (\rho V_k) = 0 \tag{1}$$

which is called the mass conservation equation, and themomentum conservation equations or generalized Navier-Stokes equations are

$$\frac{\partial g_i(\mathrm{r},t)}{\partial t} = -\nabla_k(\mathrm{r})(V_k g_i) + \nabla_j(\mathrm{r})\left(-p\delta_{ij} + \eta_{ijkl}\nabla_k(\mathrm{r})g_l\right) - \left(\delta_{ij} - \nabla_i u_j\right)\frac{\delta H}{\delta u_j(\mathrm{r},t)} - \\ \left(\nabla_i w_j\right)\frac{\delta H}{\delta w_j(\mathrm{r},t)} - \rho \nabla_i(\mathrm{r})\frac{\delta H}{\delta \rho(\mathrm{r},t)}, \quad g_j = \rho V_j \tag{2}$$

the equations of motion of phonons due to the symmetry breaking are expressed by

$$\frac{\partial u_i(\mathrm{r},t)}{\partial t} = -V_j \nabla_j(\mathrm{r})u_i - \Gamma_u \frac{\delta H}{\delta u_i(\mathrm{r},t)} + V_i, \tag{3}$$

in which $\Gamma_u$ represents phonon dissipation coefficient, and $H$ the Hamiltonian of the system, and equations of motion of phasons due to the symmetry breaking are as below

$$\frac{\partial w_i(\mathrm{r},t)}{\partial t} = -V_j \nabla_j(\mathrm{r})w_i - \Gamma_w \frac{\delta H}{\delta w_i(\mathrm{r},t)}, \tag{4}$$

in which $\Gamma_w$ represents phason dissipation coefficient. However the equation set up to now is not closed yet, because the number of field variables is greater than that of field equations. We must supplement an equation, the equation of state, i.e., the relation between fluid pressure and mass density:

$$p = f(\rho)$$

which is a difficult topic in the study of soft matter.

Wensink [31] studied the equation of stateon the columnar liquid crystals in one-dimensional case, but in our computation there are some difficulties by using it, then we [25] take some modifications as

$$p = f(\rho) = 3\frac{k_B T}{l^3 \rho_0^3}\left(\rho_0^2 \rho + \rho_0 \rho^2 + \rho^3\right) \tag{5}$$

where $\rho_0$ is the initial density, or the rest mass density, $k_B$ the Boltzmann constant, $T$ the absolute temperature, $l$ the thickness of hard disks of the columnar liquid crystals in the original paper of Wensink [31], we here take it as characteristic size of soft-matter quasicrystals, in general this is a mesocharacteristic size, $l = 1 \sim 100 nm$, and our computation shows if take $l = 8 \sim 9 nm$ the computational results are in the best accuracy.

In equations (3) and (4), the Hamiltonian $H$ is defined by

$$H = H[\Psi(\mathbf{r},t)] = \int \frac{\mathbf{g}^2}{2\rho} d^d \mathbf{r} + \int \left[ \frac{1}{2} A \left( \frac{\delta\rho}{\rho_0} \right)^2 + B \left( \frac{\delta\rho}{\rho_0} \right) \nabla \cdot \mathbf{u} \right] d^d \mathbf{r} + F_{el}$$

$$= H_{kin} + H_{density} + F_{el} \qquad (6)$$

$$F_{el} = F_u + F_w + F_{uw}, \qquad \mathbf{g} = \rho \mathbf{V}$$

in which the contributions of kinetic energy, density and elasticity terms to Hamiltonian $H$ are recorded in $H_{kin}, H_{density}, F_{el}$ respectively, $F_u, F_w, F_{uw}$ represent the phonon, phason and phonon-phason coupling strain energy densities, respectively. $\mathbf{V}$ represents the fluid velocity field, $A, B$ the constants describing density variation and density variation coupling phonon deformation, the last term of (6) represents elastic energies, which consists of phonons, phasons and phonon-phason coupling parts, for the first kind of soft-matter quasicrystals of 12-fold symmetry are as follows

$$\begin{aligned} F_u &= \int \frac{1}{2} C_{ijkl} \varepsilon_{ij} \varepsilon_{kl} d^d r \\ F_w &= \int \frac{1}{2} K_{ijkl} w_{ij} w_{kl} d^d r \\ F_{uw} &= \int \left( R_{ijkl} \varepsilon_{ij} w_{kl} + R_{klij} w_{ij} \varepsilon_{kl} \right) d^d r \end{aligned} \qquad (7)$$

## 3. The basic relationship of some physical quantities in spherical coordinate system

The spherical coordinate is shown in Fig.1

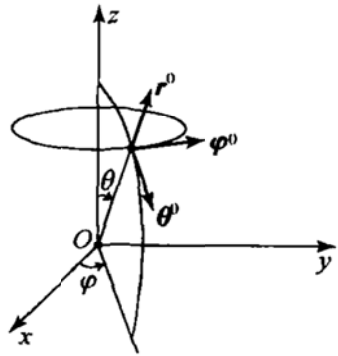

Fig.1 The spherical coordinate system and the rectilinear coordinate system

We have the phonon displacement field

$$\mathbf{u} = \mathbf{i} u_r + \mathbf{j} u_\theta + \mathbf{k} u_\varphi \qquad (8)$$

the phason displacement field

$$\mathbf{w} = \mathbf{i} w_r + \mathbf{j} w_\theta + \mathbf{k} w_\varphi \qquad (9)$$

and the fluid phonon velocity field

$$\mathbf{V} = \mathbf{i}V_r + \mathbf{j}V_\theta + \mathbf{k}V_\varphi \tag{10}$$

and the divergence operator

$$\nabla \cdot = \mathbf{i}\frac{\partial}{\partial r} + \mathbf{j}\frac{1}{r}\frac{\partial}{\partial \theta} + \mathbf{k}\frac{1}{r\sin\theta}\frac{\partial}{\partial \varphi} \tag{11}$$

and the Laplace operator

$$\nabla^2 = \frac{1}{r^2}\left(r^2\frac{\partial^2}{\partial r^2} + \frac{\partial^2}{\partial \theta^2} + \frac{1}{\sin^2\theta}\frac{\partial^2}{\partial \varphi^2} + 2r\frac{\partial}{\partial r} + \cot\theta\frac{\partial}{\partial \theta}\right) \tag{12}$$

In the spherical coordinate system, there are the phonon strain components $\varepsilon_{rr}, \varepsilon_{\theta\theta}, \varepsilon_{\varphi\varphi}, \varepsilon_{r\theta} = \varepsilon_{\theta r}$, $\varepsilon_{\theta\varphi} = \varepsilon_{\varphi\theta}, \varepsilon_{r\varphi} = \varepsilon_{\varphi r}$ with the phonon stress components $\sigma_{rr}, \sigma_{\theta\theta}, \sigma_{\varphi\varphi}, \sigma_{r\theta} = \sigma_{\theta r}, \sigma_{\theta\varphi} = \sigma_{\varphi\theta}, \sigma_{r\varphi} = \sigma_{\varphi r}$ following the generalized Hooke's law of phonons

$$\begin{aligned}
\sigma_{rr} &= C_{33}\varepsilon_{rr} + C_{13}\varepsilon_{\theta\theta} + C_{13}\varepsilon_{\varphi\varphi} \\
\sigma_{\theta\theta} &= C_{13}\varepsilon_{rr} + C_{11}\varepsilon_{\theta\theta} + C_{12}\varepsilon_{\varphi\varphi} \\
\sigma_{\varphi\varphi} &= C_{13}\varepsilon_{rr} + C_{12}\varepsilon_{\theta\theta} + C_{11}\varepsilon_{\varphi\varphi} \\
\sigma_{\theta\varphi} &= \sigma_{\varphi\theta} = 2C_{44}\varepsilon_{\theta\varphi} \\
\sigma_{\varphi r} &= \sigma_{r\varphi} = 2C_{66}\varepsilon_{r\varphi} \\
\sigma_{r\theta} &= \sigma_{\theta r} = 2C_{44}\varepsilon_{r\theta}
\end{aligned} \tag{13}$$

the phonon deformation geometry equations

$$\begin{aligned}
\varepsilon_{rr} &= \frac{\partial u_r}{\partial r}, \varepsilon_{\theta\theta} = \frac{1}{r}\frac{\partial u_\theta}{\partial \theta} + \frac{u_r}{r}, \varepsilon_{\varphi\varphi} = \frac{1}{r\sin\theta}\frac{\partial u_\varphi}{\partial \varphi} + \frac{\cot\theta}{r}u_\theta + \frac{u_r}{r} \\
\varepsilon_{r\theta} &= \varepsilon_{\theta r} = \frac{1}{2}\left(\frac{1}{r}\frac{\partial u_r}{\partial \theta} + \frac{\partial u_\theta}{\partial r} - \frac{u_\theta}{r}\right) \\
\varepsilon_{\theta\varphi} &= \varepsilon_{\varphi\theta} = \frac{1}{2}\left(\frac{1}{r\sin\theta}\frac{\partial u_\theta}{\partial \varphi} + \frac{1}{r}\frac{\partial u_\varphi}{\partial \theta} - \frac{\cot\theta}{r}u_\varphi\right) \\
\varepsilon_{r\varphi} &= \varepsilon_{\varphi r} = \frac{1}{2}\left(\frac{1}{r\sin\theta}\frac{\partial u_r}{\partial \varphi} + \frac{\partial u_\varphi}{\partial r} - \frac{u_\varphi}{r}\right)
\end{aligned} \tag{14}$$

the phason deformation geometry equations

$$\begin{aligned}
w_{rr} &= \frac{\partial w_r}{\partial r}, w_{r\theta} = \frac{1}{r}\left(\frac{\partial w_r}{\partial \theta} - w_\theta\right), w_{r\varphi} = \frac{1}{r}\left(\frac{1}{\sin\theta}\frac{\partial w_r}{\partial \varphi} - w_\varphi\right) \\
w_{\theta r} &= \frac{\partial w_\theta}{\partial r}, w_{\theta\theta} = \frac{1}{r}\left(w_r + \frac{\partial w_\theta}{\partial \theta}\right), w_{\theta\varphi} = \frac{1}{r}\left(\frac{1}{\sin\theta}\frac{\partial w_\theta}{\partial \varphi} - \cot\theta w_\varphi\right) \\
w_{\varphi r} &= \frac{\partial w_\varphi}{\partial r}, w_{\varphi\theta} = \frac{1}{r}\frac{\partial w_\varphi}{\partial \theta}, w_{\varphi\varphi} = \frac{1}{r}\left(w_r + \cot\theta w_\theta + \frac{1}{\sin\theta}\frac{\partial w_\varphi}{\partial \varphi}\right)
\end{aligned} \tag{15}$$

in the coordinate system. And the definition of the phonon strain tensors, phason strain tensor and fluid deformation rate tensors is defined by

$$\varepsilon_{ij} = \frac{1}{2}\left(\frac{\partial u_i}{\partial x_j} + \frac{\partial u_j}{\partial x_i}\right), \quad w_{ij} = \frac{\partial w_i}{\partial x_j}, \quad \dot{\xi}_{ij} = \frac{1}{2}\left(\frac{\partial V_i}{\partial x_j} + \frac{\partial V_j}{\partial x_i}\right) \tag{16}$$

in the rectilinear coordinate system respectively, which is well-known. Their form in spherical coordinate system are listed in (14), (15) and below respectively

$$\dot{\xi}_{rr} = \frac{\partial V_r}{\partial r}, \quad \dot{\xi}_{\theta\theta} = \frac{1}{r}\frac{\partial V_\theta}{\partial \theta} + \frac{V_r}{r}, \quad \dot{\xi}_{\varphi\varphi} = \frac{1}{r\sin\theta}\frac{\partial V_\varphi}{\partial \varphi} + \frac{\cot\theta}{r}V_\theta + \frac{V_r}{r}$$

$$\dot{\xi}_{r\theta} = \dot{\xi}_{\theta r} = \frac{1}{2}\left(\frac{1}{r}\frac{\partial V_r}{\partial \theta} + \frac{\partial V_\theta}{\partial r} - \frac{V_\theta}{r}\right)$$

$$\dot{\xi}_{\theta\varphi} = \dot{\xi}_{\varphi\theta} = \frac{1}{2}\left(\frac{1}{r\sin\theta}\frac{\partial V_\theta}{\partial \varphi} + \frac{1}{r}\frac{\partial V_\varphi}{\partial \theta} - \frac{\cot\theta}{r}V_\varphi\right) \tag{17}$$

$$\dot{\xi}_{r\varphi} = \dot{\xi}_{\varphi r} = \frac{1}{2}\left(\frac{1}{r\sin\theta}\frac{\partial V_r}{\partial \varphi} + \frac{\partial V_\varphi}{\partial r} - \frac{V_\varphi}{r}\right)$$

And the fluid constitute law

$$p_{ij} = -p\delta_{ij} + 2\eta\left(\dot{\xi}_{ij} - \frac{1}{3}\dot{\xi}_{kk}\delta_{ij}\right) + \eta'\dot{\xi}_{kk}\delta_{ij}$$

$$\dot{\xi}_{kk} = \dot{\xi}_{rr} + \dot{\xi}_{\theta\theta} + \dot{\xi}_{\varphi\varphi} \tag{18}$$

where we could omit the second viscosity constant $\eta'$.

## 4. Derivation of spherical coordinate form of three-dimensional generalized dynamics of soft-matter quasicrystals with 12-fold symmetry

The equations in section 2 and 3 can be summarized to get the final governing equations as follows

$$\frac{\partial \rho}{\partial t} + \nabla\cdot(\rho\mathbf{V}) = 0 \tag{19a}$$

$$\frac{\partial}{\partial t}(\rho V_r) + \frac{1}{r}\left[\left(2 + r\frac{\partial}{\partial r}\right)(V_r\rho V_r) + \left(\cot\theta + \frac{\partial}{\partial \theta}\right)(V_r\rho V_\theta) + \frac{1}{\sin\theta}\frac{\partial}{\partial \varphi}(V_r\rho V_\varphi) - (V_\theta\rho V_\theta) - (V_\varphi\rho V_\varphi)\right] =$$

$$-\frac{\partial p}{\partial r} + \eta\left[\nabla^2 V_r - \frac{2}{r^2}\left(V_r + \left(\cot\theta + \frac{\partial}{\partial \theta}\right)V_\theta + \frac{1}{\sin\theta}\frac{\partial}{\partial \varphi}V_\varphi\right)\right] + \frac{1}{3}\eta\frac{\partial}{\partial r}\nabla\cdot\mathbf{V}$$

$$+ \frac{1}{r^2}\left(C_{33}r^2\frac{\partial^2}{\partial r^2} + C_{44}\frac{\partial^2}{\partial \theta^2} + C_{44}\frac{1}{\sin^2\theta}\frac{\partial^2}{\partial \varphi^2} + 2C_{33}r\frac{\partial}{\partial r} + C_{44}\cot\theta\frac{\partial}{\partial \theta} - 2C_{11} - 2C_{12} + 2C_{13}\right)u_r$$

$$+ \frac{1}{r^2}\left((C_{13} + C_{44})r\frac{\partial^2}{\partial r\partial \theta} + (C_{13} + C_{44})r\cot\theta\frac{\partial}{\partial r} - (C_{11} + C_{12} - C_{13} + C_{44})\frac{\partial}{\partial \theta} - (C_{11} + C_{12} - C_{13} + C_{44})\cot\theta\right)u_\theta$$

$$+ \frac{1}{r^2}\left((C_{13} + C_{44})r\frac{1}{\sin\theta}\frac{\partial^2}{\partial r\partial \varphi} - (C_{11} + C_{12} - C_{13} + C_{44})\frac{1}{\sin\theta}\frac{\partial}{\partial \varphi}\right)u_\varphi - B\frac{\partial}{\partial r}(\nabla\cdot\mathbf{u}) - \frac{A-B}{\rho_0}\frac{\partial\delta\rho}{\partial r} \tag{19b}$$

$$\frac{\partial}{\partial t}(\rho V_\theta) + \frac{1}{r}\left[\left(3 + r\frac{\partial}{\partial r}\right)(V_\theta\rho V_r) + \left(\cot\theta + \frac{\partial}{\partial \theta}\right)(V_\theta\rho V_\theta) + \frac{1}{\sin\theta}\frac{\partial}{\partial \varphi}(V_\theta\rho V_\varphi) - \cot\theta(V_\varphi\rho V_\varphi)\right] =$$

$$-\frac{1}{r}\frac{\partial p}{\partial \theta}+\eta\left[\nabla^{2}V_{\theta}+\frac{1}{r^{2}}\left(2\frac{\partial}{\partial \theta}V_{r}-\frac{1}{\sin^{2}\theta}V_{\theta}-2\frac{\cos\theta}{\sin^{2}\theta}\frac{\partial}{\partial \varphi}V_{\varphi}\right)\right]+\frac{1}{3}\frac{\eta}{r}\frac{\partial}{\partial \theta}\nabla\cdot\mathbf{V}$$

$$+\frac{1}{r^{2}}\left((C_{13}+C_{44})r\frac{\partial^{2}}{\partial r\partial \theta}+(C_{11}+C_{12}+2C_{44})\frac{\partial}{\partial \theta}\right)u_{r}$$

$$+\frac{1}{r^{2}}\left(C_{44}r^{2}\frac{\partial^{2}}{\partial r^{2}}+C_{11}\frac{\partial^{2}}{\partial \theta^{2}}+C_{66}\frac{1}{\sin^{2}\theta}\frac{\partial^{2}}{\partial \varphi^{2}}+2C_{44}r\frac{\partial}{\partial r}+C_{11}\cot\theta\frac{\partial}{\partial \theta}\right)u_{\theta}$$

$$+\frac{1}{r^{2}}\left((C_{12}+C_{66})\frac{1}{\sin\theta}\frac{\partial^{2}}{\partial \theta\partial \varphi}-(C_{11}+C_{66})\frac{\cos\theta}{\sin^{2}\theta}\frac{\partial}{\partial \varphi}\right)u_{\varphi}-\frac{B}{r}\frac{\partial}{\partial \theta}(\nabla\cdot\mathbf{u})-\frac{A-B}{r\rho_{0}}\frac{\partial\delta\rho}{\partial \theta} \quad (19c)$$

$$\frac{\partial}{\partial t}(\rho V_{\varphi})+\frac{1}{r}\left[\left(3+r\frac{\partial}{\partial r}\right)(V_{\varphi}\rho V_{r})+\left(2\cot\theta+\frac{\partial}{\partial \theta}\right)(V_{\varphi}\rho V_{\theta})+\frac{1}{\sin\theta}\frac{\partial}{\partial \varphi}(V_{\varphi}\rho V_{\varphi})\right]=$$

$$-\frac{1}{r\sin\theta}\frac{\partial p}{\partial \varphi}+\eta\left[\nabla^{2}V_{\varphi}+\frac{\csc^{2}\theta}{r^{2}}\left(2\sin\theta\frac{\partial}{\partial \varphi}V_{r}+2\cos\theta\frac{\partial}{\partial \varphi}V_{\theta}-V_{\varphi}\right)\right]+\frac{1}{3}\frac{\eta}{r\sin\theta}\frac{\partial}{\partial \varphi}\nabla\cdot\mathbf{V}$$

$$+\frac{1}{r^{2}\sin\theta}\left((C_{13}+C_{44})r\frac{\partial^{2}}{\partial r\partial \varphi}+(C_{11}+C_{12}+2C_{44})\frac{\partial}{\partial \varphi}\right)u_{r}+\frac{1}{r^{2}\sin\theta}\left((C_{12}+C_{66})\frac{\partial^{2}}{\partial \theta\partial \varphi}+(C_{11}+C_{66})\cot\theta\frac{\partial}{\partial \varphi}\right)u_{\theta}$$

$$+\frac{1}{r^{2}}\left(C_{44}r^{2}\frac{\partial^{2}}{\partial r^{2}}+C_{66}\frac{\partial^{2}}{\partial \theta^{2}}+C_{11}\frac{1}{\sin^{2}\theta}\frac{\partial^{2}}{\partial \varphi^{2}}+2C_{44}r\frac{\partial}{\partial r}+C_{66}\cot\theta\frac{\partial}{\partial \theta}-2C_{44}+C_{66}-C_{66}\cot^{2}\theta\right)u_{\varphi}$$

$$-\frac{B}{r\sin\theta}\frac{\partial}{\partial \varphi}(\nabla\cdot\mathbf{u})-\frac{A-B}{r\rho_{0}\sin\theta}\frac{\partial\delta\rho}{\partial \varphi} \quad (19d)$$

$$\frac{\partial u_{r}}{\partial t}+\left(V_{r}\frac{\partial}{\partial r}+\frac{1}{r}V_{\theta}\frac{\partial}{\partial \theta}+\frac{1}{r\sin\theta}V_{\varphi}\right)u_{r}-\frac{1}{r}(V_{\theta}u_{\theta}+V_{\varphi}u_{\varphi})=V_{r}$$

$$+\frac{\Gamma_{u}}{r^{2}}\left(C_{33}r^{2}\frac{\partial^{2}}{\partial r^{2}}+C_{44}\frac{\partial^{2}}{\partial \theta^{2}}+C_{44}\frac{1}{\sin^{2}\theta}\frac{\partial^{2}}{\partial \varphi^{2}}+2C_{33}r\frac{\partial}{\partial r}+C_{44}\cot\theta\frac{\partial}{\partial \theta}-2C_{11}-2C_{12}+2C_{13}\right)u_{r}$$

$$+\frac{\Gamma_{u}}{r^{2}}\left((C_{13}+C_{44})r\frac{\partial^{2}}{\partial r\partial \theta}+(C_{13}+C_{44})r\cot\theta\frac{\partial}{\partial r}-(C_{11}+C_{12}-C_{13}+C_{44})\frac{\partial}{\partial \theta}-(C_{11}+C_{12}-C_{13}+C_{44})\cot\theta\right)u_{\theta}$$

$$+\frac{\Gamma_{u}}{r^{2}}\left((C_{13}+C_{44})r\frac{1}{\sin\theta}\frac{\partial^{2}}{\partial r\partial \varphi}-(C_{11}+C_{12}-C_{13}+C_{44})\frac{1}{\sin\theta}\frac{\partial}{\partial \varphi}\right)u_{\varphi} \quad (19e)$$

$$\frac{\partial u_{\theta}}{\partial t}+\left(V_{r}\frac{\partial}{\partial r}+\frac{1}{r}V_{\theta}\frac{\partial}{\partial \theta}+\frac{1}{r\sin\theta}V_{\varphi}\right)u_{\theta}+\frac{1}{r}(V_{\theta}u_{r}-V_{\varphi}u_{\varphi}\cot\theta)=V_{\theta}$$

$$+\frac{\Gamma_{u}}{r^{2}}\left((C_{13}+C_{44})r\frac{\partial^{2}}{\partial r\partial \theta}+(C_{11}+C_{12}+2C_{44})\frac{\partial}{\partial \theta}\right)u_{r}$$

$$+\frac{\Gamma_{u}}{r^{2}}\left(C_{44}r^{2}\frac{\partial^{2}}{\partial r^{2}}+C_{11}\frac{\partial^{2}}{\partial \theta^{2}}+C_{66}\frac{1}{\sin^{2}\theta}\frac{\partial^{2}}{\partial \varphi^{2}}+2C_{44}r\frac{\partial}{\partial r}+C_{11}\cot\theta\frac{\partial}{\partial \theta}\right)u_{\theta}$$

$$+\frac{\Gamma_{u}}{r^{2}}\left((C_{12}+C_{66})\frac{1}{\sin\theta}\frac{\partial^{2}}{\partial \theta\partial \varphi}-(C_{11}+C_{66})\frac{\cos\theta}{\sin^{2}\theta}\frac{\partial}{\partial \varphi}\right)u_{\varphi} \quad (19f)$$

$$\frac{\partial u_{\varphi}}{\partial t}+\left(V_{r}\frac{\partial}{\partial r}+\frac{1}{r}V_{\theta}\frac{\partial}{\partial \theta}+\frac{1}{r\sin\theta}V_{\varphi}\right)u_{\varphi}+\frac{1}{r}V_{\varphi}(u_{r}+u_{\theta}\cot\theta)=V_{\varphi}$$

$$+\frac{\Gamma_{u}}{r^{2}\sin\theta}\left((C_{13}+C_{44})r\frac{\partial^{2}}{\partial r\partial \varphi}+(C_{11}+C_{12}+2C_{44})\frac{\partial}{\partial \varphi}\right)u_{r}+\frac{\Gamma_{u}}{r^{2}\sin\theta}\left((C_{12}+C_{66})\frac{\partial^{2}}{\partial \theta\partial \varphi}+(C_{11}+C_{66})\cot\theta\frac{\partial}{\partial \varphi}\right)u_{\theta}$$

$$+\frac{\Gamma_{u}}{r^{2}}\left(C_{44}r^{2}\frac{\partial^{2}}{\partial r^{2}}+C_{66}\frac{\partial^{2}}{\partial \theta^{2}}+C_{11}\frac{1}{\sin^{2}\theta}\frac{\partial^{2}}{\partial \varphi^{2}}+2C_{44}r\frac{\partial}{\partial r}+C_{66}\cot\theta\frac{\partial}{\partial \theta}-2C_{44}+C_{66}-C_{66}\cot^{2}\theta\right)u_{\varphi} \quad (19g)$$

$$\frac{\partial w_{r}}{\partial t}+\left(V_{r}\frac{\partial}{\partial r}+\frac{1}{r}V_{\theta}\frac{\partial}{\partial \theta}+\frac{1}{r\sin\theta}V_{\varphi}\right)w_{r}-\frac{1}{r}(V_{\theta}w_{\theta}+V_{\varphi}w_{\varphi})=$$

$$\frac{\Gamma_w}{r^2}\{-\frac{1}{4}\sin^2\theta\left[2(5+3\cos 2\theta)K_1+(5+3\cos 2\theta+3\cos 4\varphi-3\cos 2\theta\cos 4\varphi)(K_2+K_3)-2(1+3\cos 2\theta)K_4\right]\left(1-r\frac{\partial}{\partial r}\right)$$

$$+\frac{1}{4}\sin 2\theta\left[(2+4\cos 2\theta)K_1+(1+2\cos 2\theta+3\cos 4\varphi-2\cos 2\theta\cos 4\varphi)(K_2+K_3)+4(1-\cos 2\theta)K_4\right]\frac{\partial}{\partial\theta}$$

$$+\frac{1}{2}\sin 4\varphi(\cos 2\theta-3)(K_2+K_3)\frac{\partial}{\partial\varphi}+r^2\sin^2\theta\left[K_1\sin^2\theta+(K_2+K_3)\sin^2\theta\sin^2 2\varphi+K_4\cos^2\theta\right]\frac{\partial^2}{\partial r^2}$$

$$+\sin^2\theta\left[K_1\cos^2\theta+(K_2+K_3)\cos^2\theta\sin^2 2\varphi+K_4\sin^2\theta\right]\frac{\partial^2}{\partial\theta^2}+\left[K_1+(K_2+K_3)\cos^2 2\varphi\right]\frac{\partial^2}{\partial\varphi^2}$$

$$+r\sin^2\theta\sin 4\varphi(K_2+K_3)\frac{\partial^2}{\partial r\partial\varphi}+r\sin^2\theta\sin 2\theta\left[K_1+(K_2+K_3)\sin^2 2\varphi-K_4\right]\frac{\partial^2}{\partial r\partial\theta}$$

$$+\frac{1}{2}\sin 2\theta\sin 4\varphi(K_2+K_3)\frac{\partial^2}{\partial\theta\partial\varphi}\}w_r+\frac{\Gamma_w}{r^2}\{\frac{1}{16}\cot\theta[8\cos^2\theta(3\cos 2\theta-5)K_1-(7+4\cos 2\theta-3\cos 4\theta$$

$$+42\cos 4\varphi-20\cos 2\theta\cos 4\varphi+3\cos 4\theta\cos 4\varphi)(K_2+K_3)-6(3-4\cos 2\theta+\cos 4\theta)K_4]$$

$$+\frac{1}{8}r\sin 2\theta\left[2(1+3\cos 2\theta)K_1+(1+3\cos 2\theta+7\cos 4\varphi-3\cos 2\theta\cos 4\varphi)(K_2+K_3)+12\sin^2\theta K_4\right]\frac{\partial}{\partial r}$$

$$+\frac{1}{2}[(\cos 2\theta+\cos 4\theta)K_1-2\cos^2\theta(\cos^2\theta-5\sin^2\theta\cos 4\varphi-3\cos 2\theta\cos^2 2\varphi)(K_2+K_3)+2\sin^2\theta\cos 2\theta K_4]\frac{\partial}{\partial\theta}$$

$$+\frac{1}{4}\csc\theta\sin 4\varphi(\cos 3\theta-9\cos\theta)(K_2+K_3)\frac{\partial}{\partial\varphi}+\frac{1}{2}r^2\sin 2\theta\left[K_1\sin^2\theta+(K_2+K_3)\sin^2\theta\sin^2 2\varphi+K_4\cos^2\theta\right]\frac{\partial^2}{\partial r^2}$$

$$+\frac{1}{2}\sin 2\theta\left[K_1\cos^2\theta+(K_2+K_3)\sin^2\theta\sin^2 2\varphi+K_4\sin^2\theta\right]\frac{\partial^2}{\partial\theta^2}$$

$$+\frac{1}{2}\cot\theta\left[2K_1+(1+\cos 4\varphi)(K_2+K_3)\right]\frac{\partial^2}{\partial\varphi^2}+\frac{1}{2}r\sin 2\theta\sin 4\varphi(K_2+K_3)\frac{\partial^2}{\partial r\partial\varphi}$$

$$+\frac{1}{4}r\sin^2 2\theta\left[2K_1+2\sin^2 2\varphi(K_2+K_3)-2K_4\right]\frac{\partial^2}{\partial r\partial\theta}+\cos^2\theta\sin 4\varphi(K_2+K_3)\frac{\partial^2}{\partial\theta\partial\varphi}\}w_\theta$$

$$+\frac{\Gamma_w}{r^2}(K_2+K_3)\{\frac{3}{2}\csc\theta\sin 4\varphi+\frac{1}{8}r\sin 4\varphi(\sin 3\theta-11\sin\theta)\frac{\partial}{\partial r}+\frac{1}{2}\cos\theta\sin 4\varphi(\cos 2\theta-4)\frac{\partial}{\partial\theta}$$

$$-\csc\theta(1+2\cos 4\varphi)\frac{\partial}{\partial\varphi}+\frac{1}{2}r^2\sin^3\theta\sin 4\varphi\frac{\partial^2}{\partial r^2}+\frac{1}{2}\sin\theta\cos^2\theta\sin 4\varphi\frac{\partial^2}{\partial\theta^2}-\frac{1}{2}\csc\theta\sin 4\varphi\frac{\partial^2}{\partial\varphi^2}$$

$$+r\sin\theta\cos 4\varphi\frac{\partial^2}{\partial r\partial\varphi}+r\sin^2\theta\cos\theta\sin 4\varphi\frac{\partial^2}{\partial r\partial\theta}+\cos\theta\cos 4\varphi\frac{\partial^2}{\partial\theta\partial\varphi}\}w_\varphi-2\frac{\Gamma_w}{r^2}K_1\csc\theta\frac{\partial w_\varphi}{\partial\varphi} \qquad (19\text{h})$$

$$\frac{\partial w_\theta}{\partial t}+\left(V_r\frac{\partial}{\partial r}+\frac{1}{r}V_\theta\frac{\partial}{\partial\theta}+\frac{1}{r\sin\theta}V_\varphi\right)w_\theta+\frac{1}{r}\left(V_\theta w_r-V_\varphi w_\varphi\cot\theta\right)=$$

$$\frac{\Gamma_w}{r^2}\{-\frac{1}{8}\sin 2\theta\left[2(5+3\cos 2\theta)K_1+(5+3\cos 2\theta+3\cos 4\varphi-3\cos 2\theta\cos 4\varphi)(K_2+K_3)-2(1+3\cos 2\theta)K_4\right]\left(1-r\frac{\partial}{\partial r}\right)$$

$$+\frac{1}{2}\cos^2\theta\left[(2+4\cos 2\theta)K_1+(1+2\cos 2\theta+3\cos 4\varphi-2\cos 2\theta\cos 4\varphi)(K_2+K_3)+4(1-\cos 2\theta)K_4\right]\frac{\partial}{\partial\theta}$$

$$+\frac{1}{4}\csc\theta\sin 4\varphi(\cos 3\theta-5\cos\theta)(K_2+K_3)\frac{\partial}{\partial\varphi}+\frac{1}{2}r^2\sin 2\theta\left[K_1\sin^2\theta+\sin^2\theta\sin^2 2\varphi(K_2+K_3)+K_4\cos^2\theta\right]\frac{\partial^2}{\partial r^2}$$

$$+\frac{1}{2}\sin 2\theta\left[K_1\cos^2\theta+\cos^2\theta\sin^2 2\varphi(K_2+K_3)+K_4\sin^2\theta\right]\frac{\partial^2}{\partial\theta^2}$$

$$+\frac{1}{2}\cot\theta\left[2K_1+(1+\cos 4\varphi)(K_2+K_3)\right]\frac{\partial^2}{\partial\varphi^2}+\frac{1}{2}r\sin 2\theta\sin 4\varphi(K_2+K_3)\frac{\partial^2}{\partial r\partial\varphi}$$

$$+\frac{1}{4}r\sin^2 2\theta\left[2K_1+2\sin^2 2\varphi(K_2+K_3)-2K_4\right]\frac{\partial^2}{\partial r\partial\theta}+\cos^2\theta\sin 4\varphi(K_2+K_3)\frac{\partial^2}{\partial\theta\partial\varphi}\}w_r$$

$$+\frac{\Gamma_w}{r^2}\{\frac{1}{16}\cot\theta[8\cos^2\theta(3\cos 2\theta-5)K_1-(7+4\cos 2\theta-3\cos 4\theta+42\cos 4\varphi-20\cos 2\theta\cos 4\varphi$$

$$+3\cos 4\theta\cos 4\varphi)(K_2+K_3)-6(3-4\cos 2\theta+\cos 4\theta)K_4]$$

$$+4r\cos^2\theta\Big[2(1+3\cos 2\theta)K_1+(1+3\cos 2\theta+7\cos 4\varphi-3\cos 2\theta\cos 4\varphi)(K_2+K_3)+12K_4\sin^2\theta\Big]\frac{\partial}{\partial r}$$

$$+\frac{1}{2}\cot\theta[(\cos 2\theta+\cos 4\theta)K_1-2\cos^2\theta(\cos^2\theta-5\sin^2\theta\cos 4\varphi-3\cos 2\theta\cos^2 2\varphi)(K_2+K_3)+2\sin^2\theta\cos 2\theta K_4]\frac{\partial}{\partial\theta}$$

$$+\frac{1}{2}(\cos 2\theta-5)\cot^2\theta\sin 4\varphi(K_2+K_3)\frac{\partial}{\partial\varphi}+r^2\cos^2\theta\Big[K_1\sin^2\theta+\sin^2\theta\sin^2 2\varphi(K_2+K_3)+K_4\cos^2\theta\Big]\frac{\partial^2}{\partial r^2}$$

$$+\cos^2\theta\Big[K_1\cos^2\theta+\cos^2\theta\sin^2 2\varphi(K_2+K_3)+K_4\sin^2\theta\Big]\frac{\partial^2}{\partial\theta^2}$$

$$+\frac{1}{2}\cot^2\theta\Big[2K_1+(1+\cos 4\varphi)(K_2+K_3)\Big]\frac{\partial^2}{\partial\varphi^2}+r\cos^2\theta\sin 4\varphi(K_2+K_3)\frac{\partial^2}{\partial r\partial\varphi}$$

$$+r\sin 2\theta\cos^2\theta\Big[K_1+\sin^2 2\varphi(K_2+K_3)-K_4\Big]\frac{\partial^2}{\partial r\partial\theta}+\cos^2\theta\cot\theta\sin 4\varphi(K_2+K_3)\frac{\partial^2}{\partial\theta\partial\varphi}\}w_\theta$$

$$+\frac{\Gamma_w}{r^2}(K_2+K_3)\{\frac{3}{2}\csc\theta\cot\theta\sin 4\varphi+\frac{1}{8}r\sin 4\varphi(\cos 3\theta-9\cos\theta)\frac{\partial}{\partial r}+\frac{1}{2}\cos\theta\sin 4\varphi(\cos 2\theta-4)\frac{\partial}{\partial\theta}$$

$$-\cot\theta\csc\theta(1+2\cos 4\varphi)\frac{\partial}{\partial\varphi}+\frac{1}{2}r^2\sin^2\theta\cos\theta\sin 4\varphi\frac{\partial^2}{\partial r^2}+\frac{1}{2}\cos^3\theta\sin 4\varphi\frac{\partial^2}{\partial\theta^2}$$

$$-\frac{1}{2}\csc\theta\cot\theta\sin 4\varphi\frac{\partial^2}{\partial\varphi^2}+r\cos\theta\cos 4\varphi\frac{\partial^2}{\partial r\partial\varphi}+r\cos^2\theta\sin\theta\sin 4\varphi\frac{\partial^2}{\partial r\partial\theta}$$

$$+\cos\theta\cot\theta\cos 4\varphi\frac{\partial^2}{\partial\theta\partial\varphi}\}w_\varphi-2\frac{\Gamma_w}{r^2}K_1\cot\theta\csc\theta\frac{\partial w_\varphi}{\partial\varphi} \qquad (19i)$$

$$\frac{\partial w_\varphi}{\partial t}+\left(V_r\frac{\partial}{\partial r}+\frac{1}{r}V_\theta\frac{\partial}{\partial\theta}+\frac{1}{r\sin\theta}V_\varphi\right)w_\varphi+\frac{1}{r}V_\varphi\left(w_r+w_\theta\cot\theta\right)=$$

$$\frac{\Gamma_w}{r^2}(K_2+K_3)\{\frac{3}{2}\sin^3\theta\sin 4\varphi-\frac{3}{2}r\sin^3\theta\sin 4\varphi\frac{\partial}{\partial r}+\frac{1}{2}(\cos 3\theta-2\cos\theta)\sin 4\varphi\frac{\partial}{\partial\theta}$$

$$+\frac{1}{2}\csc\theta(2-3\cos 4\varphi+\cos 2\theta\cos 4\varphi)\frac{\partial}{\partial\varphi}+\frac{1}{2}r^2\sin^3\theta\sin 4\varphi\frac{\partial^2}{\partial r^2}+\frac{1}{2}\cos^2\theta\sin\theta\sin 4\varphi\frac{\partial^2}{\partial\theta^2}$$

$$-\frac{1}{2}\csc\theta\sin 4\varphi\frac{\partial^2}{\partial\varphi^2}+r\sin\theta\cos 4\varphi\frac{\partial^2}{\partial r\partial\varphi}+r\cos\theta\sin^2\theta\sin 4\varphi\frac{\partial^2}{\partial r\partial\theta}+\cos\theta\cos 4\varphi\frac{\partial^2}{\partial\theta\partial\varphi}\}w_r+2\frac{\Gamma_w}{r^2}K_1\csc\theta\frac{\partial w_r}{\partial\varphi}$$

$$+\frac{\Gamma_w}{r^2}(K_2+K_3)\{\frac{1}{32}\csc^2\theta\sin 4\varphi[62\cos\theta-17\cos 3\theta+3\cos 5\theta]+\frac{1}{8}r\sin 4\varphi(3\cos 3\theta-11\cos\theta)\frac{\partial}{\partial r}$$

$$+\frac{1}{2}\cos\theta\cot\theta\sin 4\varphi(2\cos 2\theta-5)\frac{\partial}{\partial\theta}+\frac{1}{2}\cot\theta\csc\theta(2-5\cos 4\varphi+\cos 2\theta\cos 4\varphi)\frac{\partial}{\partial\varphi}$$

$$+\frac{1}{2}r^2\sin^2\theta\cos\theta\sin 4\varphi\frac{\partial^2}{\partial r^2}+\frac{1}{2}\cos^3\theta\sin 4\varphi\frac{\partial^2}{\partial\theta^2}-\frac{1}{2}\cot\theta\csc\theta\sin 4\varphi\frac{\partial^2}{\partial\varphi^2}+r\cos\theta\cos 4\varphi\frac{\partial^2}{\partial r\partial\varphi}$$

$$+r\sin\theta\cos^2\theta\sin 4\varphi\frac{\partial^2}{\partial r\partial\theta}+\cos\theta\cot\theta\cos 4\varphi\frac{\partial^2}{\partial\theta\partial\varphi}\}w_\theta+2\frac{\Gamma_w}{r^2}K_1\cot\theta\csc\theta\frac{\partial w_\theta}{\partial\varphi}$$

$$+\frac{\Gamma_w}{r^2}\{-\frac{1}{2}\csc^2\theta\Big[2K_1+(1-3\cos 4\varphi)(K_2+K_3)\Big]$$

$$+\frac{1}{4}r\Big[2(3+\cos 2\theta)K_1+(3+\cos 2\theta-5\cos 4\varphi+\cos 2\theta\cos 4\varphi)(K_2+K_3)+4K_4\sin^2\theta\Big]\frac{\partial}{\partial r}$$

$$+\frac{1}{2}\cot\theta\left[2K_1\cos 2\theta+(\cos 2\theta-4\cos 4\varphi+\cos 2\theta\cos 4\varphi)(K_2+K_3)+4K_4\sin^2\theta\right]\frac{\partial}{\partial\theta}$$

$$+2\csc^2\theta\sin 4\varphi(K_2+K_3)\frac{\partial}{\partial\varphi}+r^2\left[K_1\sin^2\theta+\sin^2\theta\cos^2 2\varphi(K_2+K_3)+K_4\cos^2\theta\right]\frac{\partial^2}{\partial r^2}$$

$$+\left[K_1\cos^2\theta+\cos^2\theta\cos^2 2\varphi(K_2+K_3)+K_4\sin^2\theta\right]\frac{\partial^2}{\partial\theta^2}+\csc^2\theta\left[K_1+\sin^2 2\varphi(K_2+K_3)\right]\frac{\partial^2}{\partial\varphi^2}-r\sin 4\varphi(K_2+K_3)\frac{\partial^2}{\partial r\partial\varphi}$$

$$+\frac{1}{2}r\sin 2\theta\left[2K_1+(1+\cos 4\varphi)(K_2+K_3)-2K_4\right]\frac{\partial^2}{\partial r\partial\theta}-\cot\theta\sin 4\varphi(K_2+K_3)\frac{\partial^2}{\partial\theta\partial\varphi}\}w_\varphi \qquad (19\text{j})$$

$$p=f(\rho)=3\frac{k_B T}{l^3\rho_0^3}\left(\rho_0^2\rho+\rho_0\rho^2+\rho^3\right) \qquad (19\text{k})$$

Moreover, the phason displacement component in the $z$-direction must be zero (because the $z$-axis is the periodic axis, the displacement component $w_z=0$), so that

$$w_z=w_r\cos\theta-w_\theta\sin\theta=0 \qquad (20)$$

this show the phason displacement components in $r$-direction and $\theta$-direction are not independent to each other, and the confined condition (20) makes numbers of the independent equations in (19a) to (19k) to be 10 rather than 11.

## 5. Solving steps of initial and boundary value problems of the governing equations

The mathematical structure of equations (19) is quite complicated; the exact solution of initial-boundary value problems of the equations is not available so far. We can try to construct solutions of approximate analytic version, numerical version and version of analytic-numerical joint.

### 5.1 Approximate analytic solution probe

There are possibilities to construct some approximate analytic solutions for certain special configurations, for example, for relative simpler states, e.g. in steady state or static state, for relative simpler structure geometry, e.g. the spherical symmetry or cylindrical symmetry, we can obtain the approximate analytic solutions with the hint of the classical analytic solutions in conventional fluid dynamics. The probes of ours are being carried out, but have not obtained any positive results so far.
.
### 5.2 Numerical solution, finite difference method

In the numerical methods, the finite difference approach with iterative procedure might be the best one, which are belong to systematical and modern method, the computer implementation is well down in practice. The key lies in the stability of the numerical procedure, if we choose appropriately the stability criterion the computation is stable still when the iterative steps reach a quite large number according to our practice. This computation will be successful in our practice.

### 5.3 Numerical solution, finite element method

The finite element method is also effective for the numerical solutions. In the previous derivation the Hamiltonians of the system are fully given, this provides the energy functional for the generalized variation principle of the finite element formulation, we can prove the second order variation of the functional to be non-negative. This ensures the validity of the element scheme. Comparing with the finite deference method,the finite element method presents certain flexibility, e.g. in the area near curvilinear boundary, we used curvilinear (including spherical) coordinate elements, and in other area one use rectilinear coordinate elements, this leads the computation to be greatly simplified.

**5.4 Solution based on analytic-numerical joint**

In possible cases we can try to realize some joint operation between analytic and numerical approaches, for example, we take an analytic solution as a zero-order approximate solution which can satisfy a part boundary conditions or a part of governing equations, then start to carry out alternating computation to satisfy other part of boundary conditions or other part of governing equations, the alternate calculation will be proceeded up to step in arriving at a certain species of the approximate solution.

**6. Conclusion**

The three-dimensional analysis especially in the spherical coordinate for generalized dynamics of soft-matter quasicrystals is a complicated and difficult problem, the report on the spherical coordinate equations of the generalized dynamics might be for the first time, the numerical method and computer program are being developed, the solutions will be reported in our subsequent work.

**Acknowledgement** This work is supported by the National Natural Science Foundation of China through grant 11272053. Tang Z Y thanks the support in part the National Natural Science Foundation of China through grant 11871098.